# Title: Strain-programmable van der Waals magnetic tunnel junctions

**Authors:** John Cenker[1], Dmitry Ovchinnikov[1], Harvey Yang[1], Daniel G. Chica[2], Catherine Zhu[1], Jiaqi Cai[1], Geoffrey Diederich[1,3], Zhaoyu Liu[1], Xiaoyang Zhu[2], Xavier Roy[2], Ting Cao[4], Matthew W. Daniels[5], Jiun-Haw Chu[1], Di Xiao[4,1], Xiaodong Xu[1,4,*]

[1] Department of Physics, University of Washington, Seattle, Washington 98195, USA
[2] Department of Chemistry, Columbia University, New York, NY 10027 USA
[3] Intelligence Community Postdoctoral Research Fellowship Program, University of Washington, Seattle, WA, USA
[4] Department of Materials Science and Engineering, University of Washington, Seattle, Washington 98195, USA
[5] Physical Measurement Laboratory, National Institute of Standards and Technology, Gaithersburg, MD, 20899, USA

*Corresponding author's email: xuxd@uw.edu

**Abstract:** The magnetic tunnel junction (MTJ) is a backbone device for spintronics. Realizing next generation energy efficient MTJs will require operating mechanisms beyond the standard means of applying magnetic fields or large electrical currents. Here, we demonstrate a new concept for programmable MTJ operation via strain control of the magnetic states of CrSBr, a layered antiferromagnetic semiconductor used as the tunnel barrier. Switching the CrSBr from antiferromagnetic to ferromagnetic order generates a giant tunneling magnetoresistance ratio without external magnetic field at temperatures up to $\approx 140$ K. When the static strain is set near the phase transition, applying small strain pulses leads to active flipping of layer magnetization with controlled layer number and thus magnetoresistance states. Further, finely adjusting the static strain to a critical value turns on stochastic switching between metastable states, with a strain-tunable sigmoidal response curve akin to the stochastic binary neuron. Our results highlight the potential of strain-programmable van der Waals MTJs towards spintronic applications, such as magnetic memory, random number generation, and probabilistic and neuromorphic computing.

**Main Text:**

The control and readout of discrete magnetic states lies at the foundation of the fields of spintronics and modern information storage[1-4]. Standard spintronic devices utilize the spin filtering phenomenon, where spin-selective transport processes, such as electron tunneling through magnetic layers, create spin polarization and magnetoresistance[5-10]. Controlling the energetics and stability of the magnets in such devices, known as magnetic tunnel junctions (MTJ), has enabled many important technological advancements. For instance, switching the orientation of the magnets from anti-parallel (AP) to parallel (P) in stable MTJs results in large changes to the tunneling magnetoresistance (TMR). This behavior is the conceptual basis for magnetic random-access memory (MRAM). On the other hand, when the magnetic layers are thinned so that the energy difference between P and AP states is small, the magnetic order becomes unstable and stochastic switching between the two states is observed[11-15]. Such stochastic MTJs can serve as probabilistic bits (p-bits), the fundamental building blocks for the emerging fields of probabilistic and neuromorphic computing[11,16]. Despite the great successes of conventional MTJs in both conventional and probabilistic computing schemes, writing the magnetic memory bits in current MRAM schemes tends to rely on energy-intensive means such as the application of large magnetic fields or currents[17]. Moreover, since the stability of the MTJ is fixed by the growth thickness, it is difficult to switch from stable MRAM operation to unstable p-bit functionality in the same device.

The recent discovery[18] of a reversible strain-induced magnetic phase transition in the air stable A-type layered antiferromagnetic (AFM) semiconductor CrSBr could offer both a new material platform and operating principle for controlling atomically thin MTJs. The A-type AFM configuration consists of van der Waals (vdW) layers with intralayer ferromagnetic (FM) order and interlayer AFM coupling along the stacking direction, forming intrinsic spin filters that can generate exceptionally large TMR[19-22]. These previous works have demonstrated that applying an external magnetic field to A-type antiferromagnets with weak interlayer exchange switches the magnetic state from the AFM, high resistance configuration to intermediate states with layer-dependent interlayer coupling, and then finally to a low resistance, field-induced FM state (Fig. 1a). In comparison to the previously studied devices which require continuous application of magnetic field to control the magnetic states, strain could provide an exceptionally energy-efficient operating mechanism as it requires essentially no current. Moreover, the fine, continuous, and reversible tuning of the interlayer exchange could enable unprecedented control of the layer-dependent magnetic structure.

Here, we demonstrate a strain-controlled vdW MTJ with programmable magneto-resistance states and stochastic switching, charting a path towards new memory and computing technologies. The schematic for our strain device is shown in Figure 1b. The vdW MTJ heterostructure is composed of a CrSBr tunnel barrier sandwiched between two narrow graphite contacts. The whole MTJ is fixed to a stretchable polyimide substrate by a gold clamp with a small (≈ 5 µm) window around the junction (Methods). This design ensures a highly efficient strain transfer when the polyimide substrate is stretched by a home-built piezoelectric strain cell[18,23], while also allowing for optical spectroscopy measurements of the junction region. The strain is applied along the crystallographic *a* axis for consistency with previous experiments[18]. The data in the main text is taken on a MTJ with an ≈ 11 nm tunnel barrier, but the technique is compatible with CrSBr flakes of any thickness.

Figure 1c shows the tunneling magnetoresistance as a function of magnetic field ($\mu_0 H$) applied along the *c* axis. In the low strain condition with piezo voltage ($V_p$) of -5 V, CrSBr is in the AFM state at $\mu_0 H$ = 0 T. As $|\mu_0 H|$ increases, the spins cant from the AFM configuration, gradually increasing the conductivity of the MTJ until it reaches the field-induced FM state with $|\mu_0 H|$ > 1 T. This behavior is consistent with the in-plane A-type layered AFM order in CrSBr[24]. We note that the saturating field is lower than standard exfoliated CrSBr samples due to a built-in strain which we determine to be ≈ 0.9 % from the Raman spectra (Methods, and Extended Data Fig. 1). Using the difference in resistance between the FM ($R_p$) and AFM ($R_{ap}$) states, we find the tunneling magnetoresistance ratio to be TMR (%) = $\frac{R_{AP} - R_p}{R_p}$ ≈ 3100 %, on par with other 2D A-type AFM tunnel junctions[19-21], albeit at much higher operating temperature.

**Strain Switching MTJ**

When the piezo voltage is increased, the TMR decreases dramatically (Fig. 2a). Furthermore, the shape of the tunneling magnetoresistance curves evolves from a giant, purely negative magnetoresistance (i.e., decreasing resistance with increasing field) at low strain to small positive MR at high strain (Extended Data Fig. 2), with complex, hysteretic behavior in between, e.g., the curve at 5 V in Fig. 2A. The large decrease in TMR and switching from negative to positive magnetoresistance implies that the interlayer magnetic coupling is switched from AFM to FM at large strain. This picture is confirmed by comparison of the strain-dependent photoluminescence (PL) with the magnetoresistance. The PL shows the characteristic red-shift from the strain induced AFM to FM phase transition, as demonstrated in a previous report[18], which is concurrent with the large changes in tunneling magnetoresistance (Figs. 2b-c). The close correspondence between the magneto-PL and tunneling magnetoresistance is a consequence of the coupling of spin and charge in magnetic semiconductors, which forbids or allows interlayer electronic hybridization and tunneling in the AFM and FM states, respectively.

In the low-strain state, the A-type AFM structure creates tunnel barriers composed of spin filters with alternating spin orientation. In the FM state, however, the tunnel barrier is uniform for all layers, i.e., all spin filters are aligned in the same direction. As a result, applying a saturating magnetic field at low strains strongly enhances the tunneling current with respect to the AFM state (top panel, Fig. 2d). At high strains, however, there is little difference between the zero- and high magnetic field tunneling behavior, as expected for a FM tunnel barrier[25] (Fig. 2d, bottom). The combination of optical and tunneling measurements unambiguously prove that the strain-induced AFM to FM phase transition is the cause of the large tunneling magnetoresistance switching, excluding trivial origins such as contact failure during the straining process.

We realized strain switching of the MTJ at zero magnetic field. Figure 3a shows the tunneling resistance as the piezo voltage is continually increased. At around 5 V, the sample experiences a switch from AFM to FM states accompanied by a sharp drop in resistance. This strain-induced phase transition generates a TMR ratio of ≈ 2700 %, comparable to the field-induced TMR in the AFM state. When the tension is released, the resistance recovers to its original value. The observed hysteresis between up and down strain sweeps is likely due to a combination of the piezo stack hysteresis and hysteresis in the first-order magnetic phase transition itself. This switching operation is robust over many cycles, with no obvious slipping or degradation over the entire measurement (> 50 strain sweeps).

The strain-switching operation of the MTJ persists to much higher temperature than other 2D MTJs[19-22,25-27]. Figure 3b shows tunneling magnetoresistance vs strain cycles at select temperatures. At higher temperatures, the transition between low and high tunneling magnetoresistance states becomes broader, but a large strain switching ratio is maintained. As shown in Fig. 3c, the zero-field strain-induced TMR exceeds 10,000 % at 30 K and remains above 100 % up to ≈ 140 K. Interestingly, a dome of positive magnetoresistance as a function of field can still be induced by a large strain at 155 K, well above the Neel temperature of 132 K reported in previous studies[24,28,29] (Fig. 3d). A likely explanation is that the enhancement of the interlayer FM exchange induces a long-range ordering of the previously reported intermediate FM (iFM) phase where the individual layers are ferromagnetically ordered, but the interlayer coupling remains paramagnetic[29].

**Strain programmable layer-dependent magnetism**

An intriguing feature of the strain-dependent TMR sweeps is that there are multiple resistance jumps during the AFM-FM phase transition, indicating the formation of multiple magnetic domains in the junction area of about 500 x 500 nm$^2$. These domains are also evident from the complex, hysteretic behavior observed in the field dependent TMR measurements (Fig. 2a, 5 V). Similar magnetic domain behavior is observed in both the nanoscale junction region and across several microns of the sample in magneto-PL (Extended Data Fig. 3). These results suggest the formation of vertical instead of lateral magnetic domains during the phase transition. The domains may arise from small vertical strain gradients. Thus, near the critical strain of the magnetic phase transition, the interlayer coupling can be FM for some layers and AFM for others. These layer-wise magnetic domains could serve as individual magnetic memory states which can be precisely manipulated by strain.

To explore active control of layer magnetization flipping, we set the static strain near the phase transition and then apply strain pulses with a small and controllable amplitude $V_{PAC}$ (see Figure 4a inset). Figure 4a shows the tunneling current over time as $V_{PAC}$ is increased from 5 mV to 0.25 V. As the pulse reaches an amplitude of ≈ 24 mV, corresponding to a strain of only ≈ 0.0008 %, the amplitude of tunneling current pulses jumps into a distinctly stable state (left-most purple arrow in Fig. 4a). This indicates the MTJ switches between two magnetization states with the strain pulse actively flipping the magnetization direction of individual layers. Calculating the gauge factor, $GF = \frac{\frac{\Delta R}{R}}{\varepsilon}$, gives an exceptionally large value of ≈ 3500, among the largest value reported in any system[30,31].

By increasing the magnitude of the strain pulse, the number of layers whose magnetization can be flipped also increases. This is evidenced by the additional distinct jumps in tunneling current with increasing pulse amplitude (purple arrows in Fig. 4a). With a large enough strain pulse, the static state current abruptly increases, indicating a change in the static magnetic configuration. This behavior is completely different than what is observed in the purely FM or purely AFM states, where increasing strain pulse magnitude only produces small, continuous changes at a gauge factor three orders of magnitude smaller, and with no change in the static current (Extended Data Fig. 4). Therefore, we conclude that the strain pulse switching observed in Fig. 4a arises from changing the vertical domain structure of the mixed magnetic states. These results demonstrate that multiple individual magnetic domains, including the static magnetic state, can be controlled by applying extremely small strain pulses.

**Stochastic domain switching**

The demonstrated ability to switch the layer-dependent magnetization suggests that strain can tune the MTJ into a regime where the AFM and FM interlayer couplings are extremely close in energy. Starting from a stable magnetic domain structure, we increase the static strain, $V_{PDC}$, by 14 mV, as indicated by the red arrow in top panel of Fig. 4b. In such a condition, the tunneling current proceeds to fluctuate between two values (Fig. 4b, bottom). By decreasing the piezo voltage back to the original value (blue arrow in Fig. 4b, top), the tunneling current returns to a stable value. The current fluctuations can be reliably turned on and off, as demonstrated. To our knowledge, this is the first realization of p-bit type operation using a vdW MTJ. This functionality is enabled by the unique ability of strain to finely and continuously tune the energy barrier between parallel and anti-parallel spin configurations, enabling in-situ switching from stable, MRAM type to stochastic, p-bit type domains (Fig. 4c).

By defining the lower current state as a 0 and the higher current state as a 1, we can convert the data to a binary sequence and analyze how the statistics of the domain switching respond to external control knobs, i.e. applied bias voltage and strain. We find that increasing the bias voltage applied to the tunnel junction leads to a large increase in the switching rate (Fig. 4d). Intriguingly, no switching is observed when a current of similar magnitude flows in the opposite direction (Extended Data Fig. 5). This bias-polarity dependence implies that heating is not the origin of the increased switching rate. Instead, the data suggests that the sample has an asymmetric vertical magnetic domain structure which creates a difference in spin polarization and thus spin transfer torque effects when the current is passed in opposite directions[12] (Extended Data Fig. 5). Whether such an asymmetric domain structure can give rise to exchange bias[32], magnetic ratchet effect[33], and other spintronics physics within a single crystal is a fascinating direction for future studies.

The relatively high Neel temperature ($T_N$ =132 K) of CrSBr in comparison to other 2D A-type AFMs creates opportunities for potential device applications operating above liquid nitrogen temperature. Figure 4e shows the response function ($\rho$) of the MTJ as a function of the static piezo voltage with a starting value near the strain-induced phase transition at 85 K. The response function is calculated by converting the MTJ output to a binary sequence and calculating the average over the entire time window. Therefore, a response function value of 0 or 1 indicates a stable magnetic domain, while a value of 0.5 indicates equal fluctuations between the two stable states. The ability to finely tune the response function should enable both random number generation at $\rho = 0.5$ and a biased Bernoulli sequence at higher or lower values, which can be important for applications dealing with Ising and probabilistic computing[12]. We further note that the applied bias voltage may also be used to tune the response function by increasing or decreasing the switching rate, potentially providing fine control near the edges of the sigmoidal curve, while also enabling interaction between multiple p-bits. In principle, the two independent control parameters (strain and bias voltage) could also offer independent tuning of the effective temperature and energy landscape of the p-bit, thereby allowing direct stochastic annealing of a p-bit system. Such a scheme could significantly reduce the circuit complexity required to realize a large-scale analog p-bit annealer, though additional study is needed to establish the full mapping between our two-dimensional voltage landscape and the statistical mechanical state space of the p-bit dynamics.

To test the stochasticity of our device, we analyze the switching data taken when $\rho \approx 0.5$, generating a binary sequence with near equal 1s and 0s, as shown in Figs. 4f-g. Since the lock-in detection scheme reads the current much faster than the domain switching rate, we sample the raw data at a frequency which is slower than the calculated switching rate to prevent non-random runs of 1s and 0s (see discussion in Supplementary Information). We tested the data using the NIST test suite (Fig. 4g) and by analyzing the rise and dwell time of the switching events, which shows that the device spends equal amounts of time in the 0 and 1 state within the experimental error (Supplementary Information). These analyses combined with their physical origin strongly suggests that the metastable states switch stochastically, thereby acting as a random number generator.

In conclusion, we have demonstrated that strained single crystal CrSBr offers a powerful platform for realizing zero-field programmable spintronic devices down to the atomically thin limit (Extended Data Fig. 6). Due to the versatile nature of vdW heterostructures, our results create a new path for various other programmable 2D quantum devices. For instance, replacing the graphite contacts with superconducting ones could enable field-free control of magnetic Josephson junctions[34-37] and superconducting diode effects[38-40]. Moreover, the ability to switch the layer-dependent magnetization and vertical magnetic domain structure creates unprecedented opportunities to precisely vary the length of the FM and AFM tunnel barriers in-situ without significantly changing the overall thickness of the insulating CrSBr barrier layer. This capability could provide a new platform for exploring exotic phenomena that have been proposed in superconductor/ferromagnetic junctions with inhomogeneous magnetization such as spin triplet correlations. More generally, our clamping and strain technique greatly expands the accessible strain range for cryogenic transport experiments on 2D devices, which could enable exciting discoveries on the emergent quantum phenomena in vdW heterostructures including moiré systems.

**Methods**

**Device fabrication and strain application**

To prepare the strain substrate, we first cut transparent 20 µm thick polyimide into strips and epoxied them onto 2D flexure sample plates produced by Razorbill instruments[†] using Stycast 2850 FT epoxy. The distance between the edge of the epoxy on either side of the gap was less than 200 µm to enable large strains.

Bulk CrSBr crystals were grown by the same method detailed previously[28]. The bulk CrSBr and graphite crystals were exfoliated onto PDMS substrates using standard methods and thin (~ 10 nm) flakes were identified by optical contrast. The MTJs were then assembled through a dry transfer technique with a stamp consisting of a polypropylene carbonate (PPC) film spin coated onto a polydimethylsiloxane (PDMS) cylinder. The flakes were picked up in the following order before being deposited onto the polyimide substrate: top graphite, CrSBr, bottom graphite. The long axis of the CrSBr flake was aligned with the strain axis for consistency with the previous studies[18].

After depositing the MTJ heterostructure, the window clamping pattern and electrical contacts to the two graphite contacts were fabricated using standard electron beam lithography techniques with a metal thickness of 7 and 70 nm Cr and Au, respectively. Then, the sample plate was screwed into the same symmetric three-piezo strain cell used previously[18,23] for strain experiments on bulk crystals and our previous experiments on strained CrSBr.

To calibrate the strain during the experiment, we used the same Raman shift rate of the mode near ~ 346 cm$^{-1}$ that we determined in the previous study[2]. We found that there was a rather large built-in strain of ~ 0.9 %, which is consistent with the small saturating field in the out-of-plane direction.

---

[†] Certain commercial processes and software are identified in this article to foster understanding. Such identification does not imply recommendation or endorsement by the National Institute of Standards and Technology, nor does it imply that the processes and software identified are necessarily the best available for the purpose.

The observation that the strain-induced phase transition occurs at negative piezo voltages at lower temperature is consistent with a thermally induced built-in strain which increases with cooling.

**Optical measurements:**

Optical measurements were performed using a backscattering geometry in a closed-cycle helium cryostat (Opticool by Quantum Design) with a nominal sample temperature of 60 K. An objective lens focused 632.8 nm light from a He/Ne laser to a spot size of ~ 1 µm. For Raman measurements, a laser power of 200 µW was used and the collected signal was dispersed using a 1800 mm$^{-1}$ groove-density grating and detected by an LN-cooled charge-coupled device (CCD) with an integration time of 210 seconds. BragGrate$^{TM}$ notch filters were used to filter out Rayleigh scattering down to ~10 cm$^{-1}$. A roughly linear background originating from weak polyimide photoluminescence was subtracted to increase the accuracy of the fitting results. For photoluminescence measurements, we used a laser power of 50 µW focused by the same objective. The collected light was dispersed by a 600 mm$^{-1}$ groove-density grating and detected by the same CCD with a 20 second integration time.

**Transport measurements:**

Except for the data presented in Extended Data Fig. 6, the transport measurements were performed in the same measurement conditions (Opticool by Quantum Design) as the optical ones, enabling direct comparison between the observed phenomena. The data shown in Figures 1-3 and 4b are taken using standard two terminal DC measurements with a Keithley 2450, while the rest of the data in Figure 4 are taken using AC detection with a DC offset voltage applied by a Zurich Instruments HF2 lock-in amplifier. The current was amplified by a current preamplifier (DL Instruments; Model 1211) with a sensitivity of 1 V/10$^{-6}$ A. For the switching data used in Fig. 4E-F and the stochasticity analysis, a time constant of 5.082 ms with a fourth-order filter was used, which was found to give the best time resolution while maintaining a high signal to noise ratio. The current was amplified by a current preamplifier (DL Instruments; Model 1211) with a sensitivity of 1 V/10$^{-6}$ A.

The 6L device in Extended Data Fig. 6 was measured in a PPMS DynaCool cryostat by Quantum Design. The data in Fig. S6a-c were taken using the same AC detection scheme, but with an SR860 lock-in amplifier. The switching data in Fig. S6d-e were obtained using a constant current measurement scheme, which was achieved by putting a 100 MΩ resistor in series with the device. The resistance signal was then pre-amplified by the differential-ended mode of SR560 with 20 times amplification.

**Acknowledgements:** We thank Xuetao Ma and Yen-Cheng Kung for fabrication advice, G.C. Adam, W.A. Borders, and J. J. Mcclelland for proofreading the paper, and John Stroud and Heonjoon Park for their help during the initial stages of the project. The strain controlled optical measurement is mainly supported by DE-SC0018171. The strain-controlled tunneling experiment is mainly supported by Air Force Office of Scientific Research (AFOSR) Multidisciplinary University Research Initiative (MURI) program, grant no. FA9550- 19-1-0390. CrSBr crystal synthesis is supported by the Center on Programmable Quantum Materials, an Energy Frontier Research Center funded by the U.S. Department of Energy (DOE), Office of Science, Basic Energy Sciences (BES), under award DE-SC0019443. DGC is supported by the Columbia MRSEC on Precision-Assembled Quantum Materials (PAQM) (DMR-2011738). XX acknowledges support from the State of Washington funded Clean Energy Institute and from the Boeing Distinguished Professorship in Physics. JC acknowledges the Graduate Fellowship from Clean Energy Institute funded by the State of Washington. ZL and JHC acknowledge the support of the David and Lucile Packard Foundation. This research was supported by an appointment to the Intelligence Community Postdoctoral Research Fellowship Program at University of Washington, administered by Oak Ridge Institute for Science and Education through an interagency agreement between the U.S. Department of Energy and the Office of the Director of National Intelligence.

**Author contributions:** XX and John C conceived the project. John C performed the optical and transport measurements with help from Jiaqi C and GD. DO supervised transport measurements and contributed to fabrication development. John C fabricated the samples with assistance from HY and ZL. John C, DO, TC, JHC, DX, and XX analyzed the data and interpreted the results. TC, MWD and DX provided theoretical support. DGC grew the CrSBr crystals with supervision from XR and XYZ. John C and XX wrote the manuscript with input from all authors. All authors discussed the results.

**Competing interests:** John C and XX have applied for a patent based on this work.

**Data availability:** The datasets generated during and/or analyzed during this study are available from the corresponding author upon reasonable request.

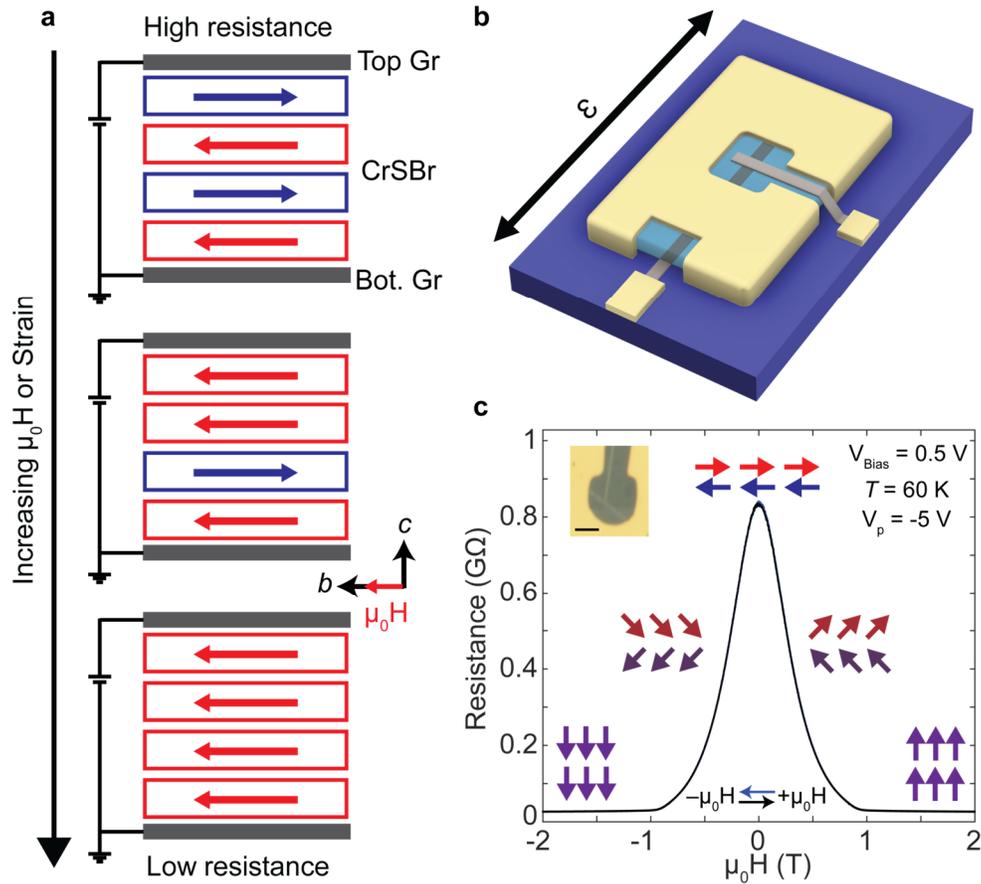

**Figure 1 | Straintronic van der Waals magnetic tunnel junction. a,** Schematic of the magnetic state evolution of the CrSBr tunnel barrier with the application of either magnetic fields along the easy *b* axis or in-plane uniaxial strain. The changing magnetic configuration creates different resistance states when bias is applied between the graphite contacts (grey). The red and blue arrows denote the spin direction within each layer. **b,** Schematic of straintronic MTJ consisting of graphite contacts sandwiching a CrSBr tunnel barrier (blue). The whole device is fixed by gold clamps to a flexible polyimide substrate (purple) which is then strained. **c,** Magnetic field dependence of a MTJ using an ≈ 11 nm CrSBr tunnel barrier (optical image inset, scale bar 3 µm) at a temperature of 60K. The device is slightly strained but remains in the AFM state at zero magnetic field. Magnetic field is applied along the hard *c* axis, leading to spin canting (inset arrows).

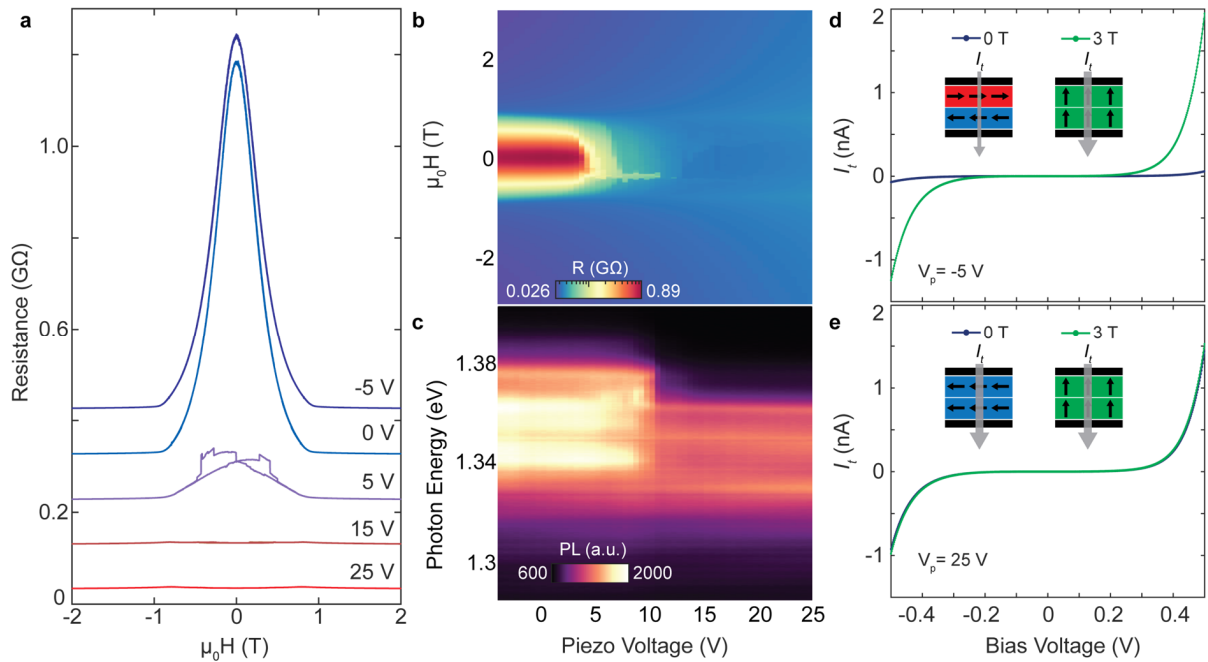

**Figure 2 | Strain switchable magnetic tunnel junctions. a,** Magnetoresistance sweeps at select piezo voltages with a fixed bias voltage across the MTJ of $V_B = 0.5$ V. The sweeps are offset for clarity. **b,** Full strain-dependent tunneling magnetoresistance with the magnetic field swept from positive to negative. **c,** Strain dependent photoluminescence intensity plot. The beam spot was kept fixed on the junction region while the strain was continuously swept. **d-e,** Bias dependent tunneling current with magnetic fields of 0 T (blue) and 3 T (green) applied in the low strain (**d**) and high strain (**e**) states. The magnetic state for each curve is depicted in the inset. All measurements were performed at a temperature of 60 K.

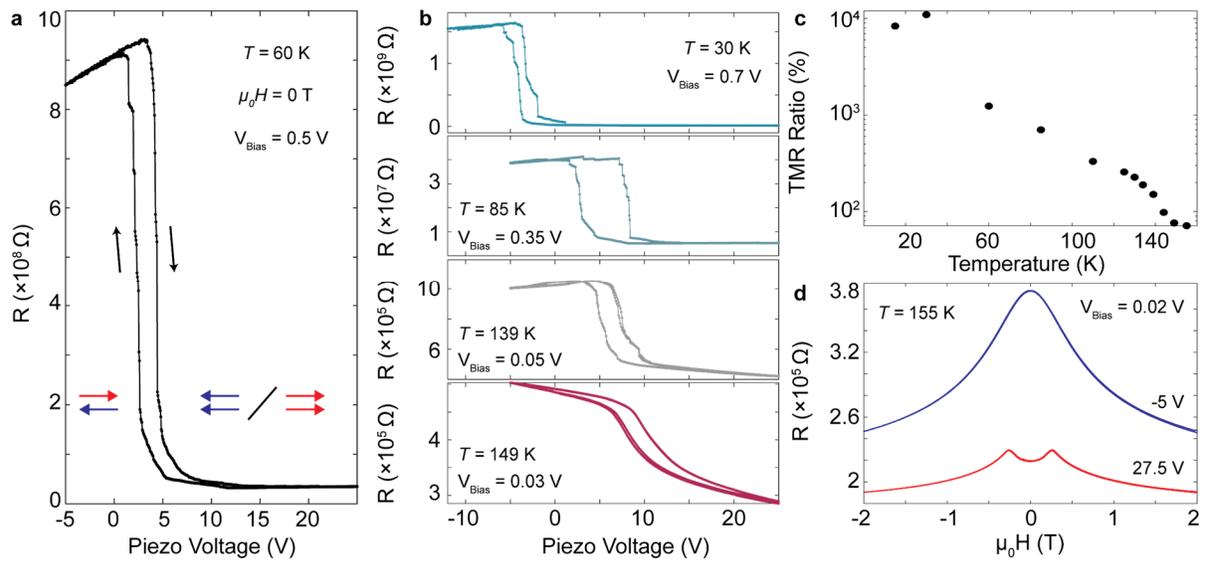

**Figure 3 | Temperature dependent zero-field tunneling resistance switching. a,** Tunneling resistance as a function of piezo voltage. A large TMR change of ≈ 2700 % is observed between the low and high strain states at 60 K. The change in magnetic state from AFM to FM interlayer coupling is depicted by the inset spin diagram. **b,** Piezo-voltage-dependent tunneling resistance at select temperatures from 30 K to 149 K. **v**, Temperature dependence of the tunneling magnetoresistance ratio, defined as TMR (%) = $\frac{R_{ap}-R_p}{R_p} \times 100$. **d,** Magnetic-field dependent tunneling resistance at 155 K in the low strain (blue) and high strain (red) states.

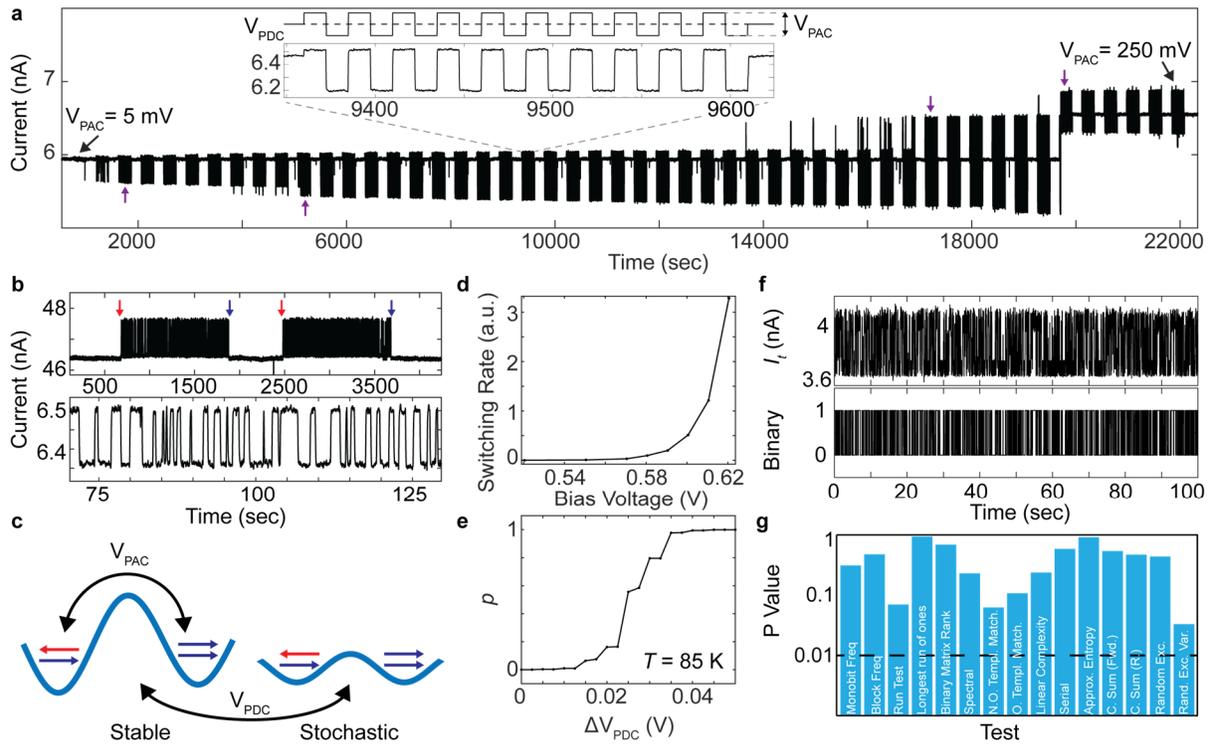

**Figure 4 | Strain control of multiple stable and stochastic layer-dependent magnetic domains. a,** Tunneling current over time as strain pulses of increasing amplitude are applied. The inset shows the measurement scheme: a small pulse of amplitude $V_{PAC}$ is applied on top of a static piezo voltage $V_{PDC}$. The system is initialized by slowly increasing $V_{PDC}$ until the magnetic phase transition starts to occur. As the pulse amplitude increases, the current switching stabilizes into discrete states (denoted by the purple arrows). Additionally, the resting current, i.e. ground state, can be changed by a sufficiently large pulse. **b,** Tunneling current over time as the static piezo voltage, $V_{PDC}$ is increased (red arrow) and then decreased (blue arrow) by .014 V. No strain pulse is applied. Bottom: Finer time resolution data of the domain fluctuations observed in the top panel. **c,** Schematic of strain tuning between magnetic domains. A sufficiently high pulse, $V_{PAC}$, will flip between AFM and FM domains (left). The fine adjustment of the static strain lowers the energy difference between AFM and FM domains, creating a metastable state with stochastic domain switching (right). **d,** Bias dependence of the switching rate in the metastable state. The piezo voltage is kept constant during the measurement. Data from panels A-D are taken at 60 K. **e,** Response function of a sensitive magnetic domain as a function of static piezo voltage at a temperature of 85 K. A value of either 0 or 1 indicates a stable domain. **f,** Tunneling current (top) and converted binary sequence (bottom) over time when the response function is near 0.5, indicating equal amount of fluctuations between the parallel and antiparallel configuration. **g,** P-values returned by the NIST random number test suite applied to the binary sequence from **f**. The black dashed line indicates a p-value of .01, the threshold for passing the specific test. The sampling time was .1760 seconds (see Supplementary Information).

# Extended Data for

# Strain-programmable van der Waals magnetic tunnel junctions

**Authors:** John Cenker[1], Dmitry Ovchinnikov[1], Harvey Yang[1], Daniel G. Chica[2], Catherine Zhu[1], Jiaqi Cai[1], Geoffrey Diederich[1,3], Zhaoyu Liu[1], Xiaoyang Zhu[2], Xavier Roy[2], Ting Cao[4], Matthew W. Daniels[5], Jiun-Haw Chu[1], Di Xiao[4,1], Xiaodong Xu[1,4,*]

[1] Department of Physics, University of Washington, Seattle, Washington 98195, USA
[2] Department of Chemistry, Columbia University, New York, NY 10027 USA
[3] Intelligence Community Postdoctoral Research Fellowship Program, University of Washington, Seattle, WA, USA
[4] Department of Materials Science and Engineering, University of Washington, Seattle, Washington 98195, USA
[5] Physical Measurement Laboratory, National Institute of Standards and Technology, Gaithersburg, MD, 20899, USA

*Correspondence to: xuxd@uw.edu

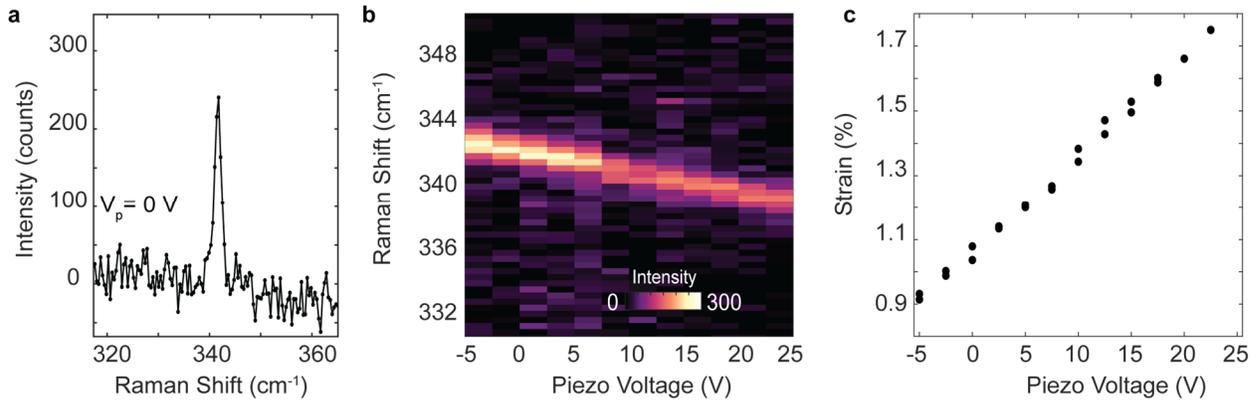

**Extended Data Fig. 1 | Calibration of strain through Raman spectroscopy. a,** Raman scattering from the $P_3$ phonon taken on the tunnel junction region at a piezo voltage of 0 V. A linear background originating from the polyimide photoluminescence is subtracted. The narrow linewidth indicates a homogenous strain. **b,** Raman intensity plot as a function of piezo voltage. The beamspot is kept on the junction as the piezo voltage is continually increased. **c,** Measured strain as a function of the applied voltage to the strain cell. The strain is calculated by fitting the data from **b** with Lorentzian fits and then comparing the peak position to the unstrained value of 346 cm$^{-1}$ using a strain shift rate of 4.2 cm$^{-1}$/% as reported in previous studies. We found that there was a built-in strain of ~ 0.9 % at the lowest piezo voltage used at this temperature.
22

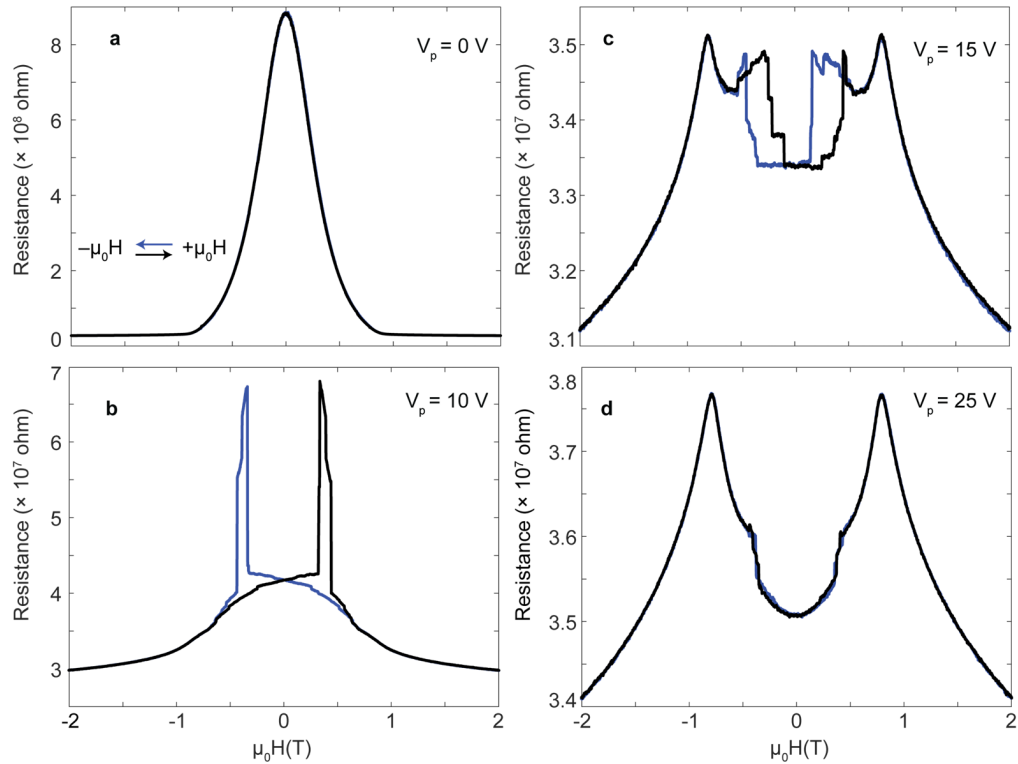

**Extended Data Fig. 2 | Magnetoresistance sweeps at select piezo voltages. a-d,** Magnetoresistance sweeps as the field is swept down (blue) and up (black) at select piezo voltages through the strain-induced layered magnetization flipping. At low strain (**a**), large negative magnetoresistance is observed, consistent with AFM order, while small positive magnetoresistance is observed in the high strain induced FM state (**d**). In between, complex and hysteretic magnetic domain behavior is observed.

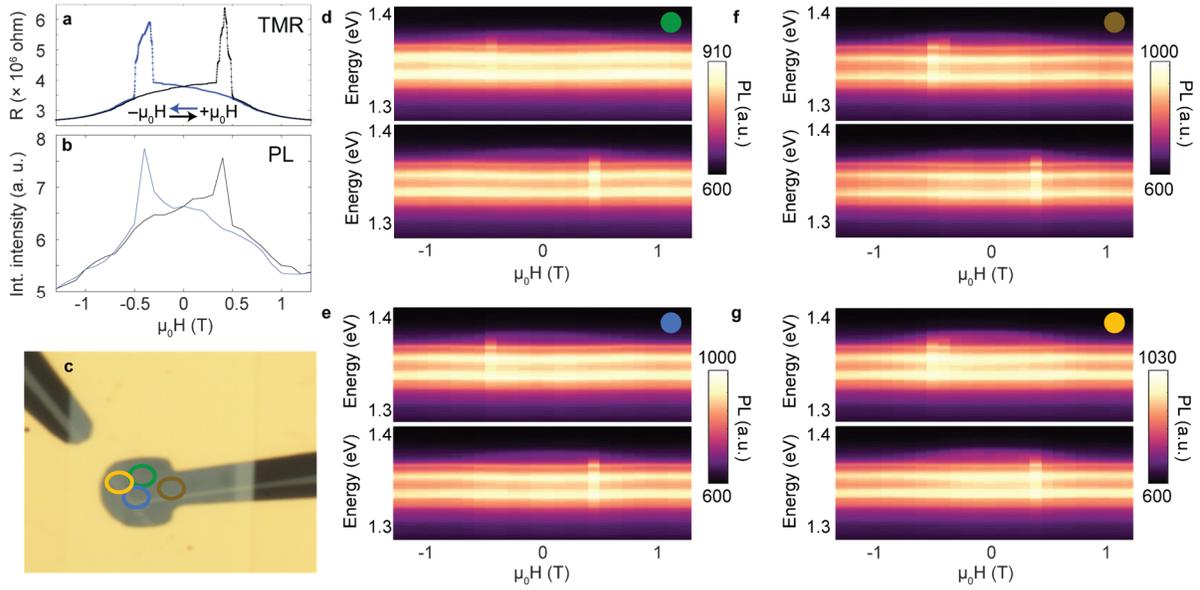

**Extended Data Fig. 3 | Magneto-photoluminescence mapping of magnetic domains. a-b,** Comparison of tunneling magnetoresistance (**a**) and integrated intensity from magneto-photoluminescence (PL) (**b**) measurements at the same piezo voltage. The correlation of the curves highlights the connection of the interlayer magnetic coupling to both electronic tunneling and exciton luminescence. **c,** Optical image of the device with different spots labeled by different colors. **d-g,** Magneto-PL sweeps at each of the spots labeled in (**c**). The similarities between spots separated by several microns indicates the presence of vertical, rather than lateral, magnetic domains.

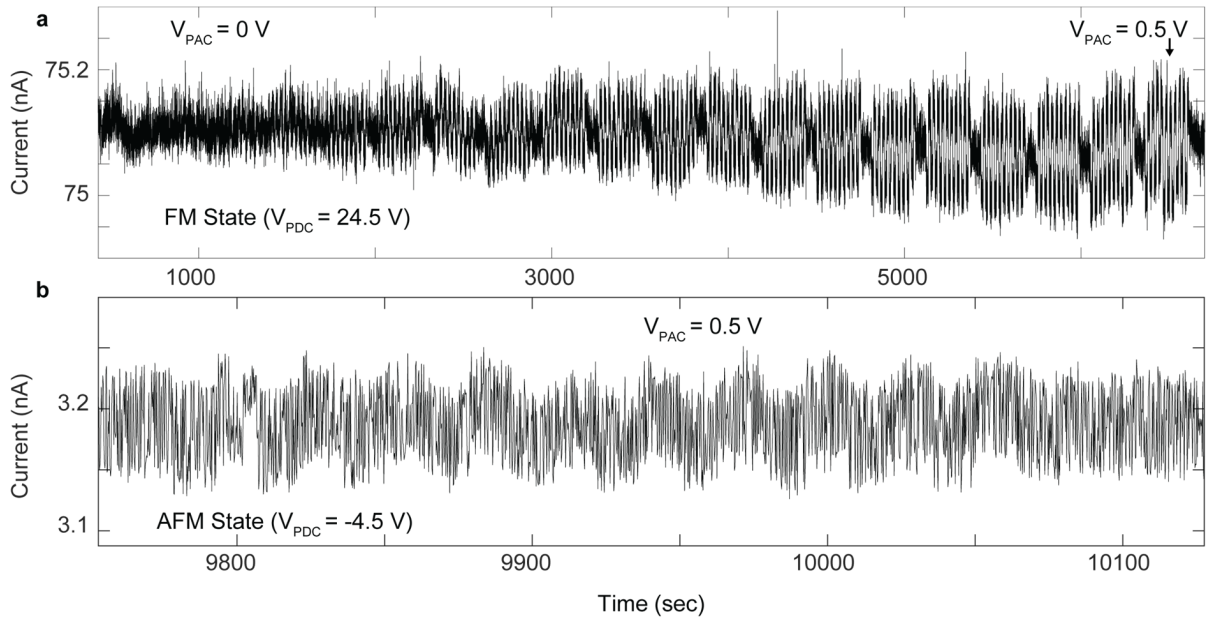

**Extended Data Fig. 4 | Strain pulse data in the purely FM and AFM states. a,** Strain pulse amplitude dependence in the purely FM state. As $V_{PAC}$ is increased from 0 to 0.5 V, a continuous change in the current is observed. The calculated gauge factor is ~ 5. **b,** Change in tunneling current over time as a strain pulse of 0.5 V is applied in the AFM state. Due to the very large resistance, the effect of pulses with smaller amplitude cannot be resolved. A gauge factor of ~ 30 is calculated, but with a large uncertainty due to the high resistance in the AFM state. No changes to the static current are observed in either FM or AFM states.

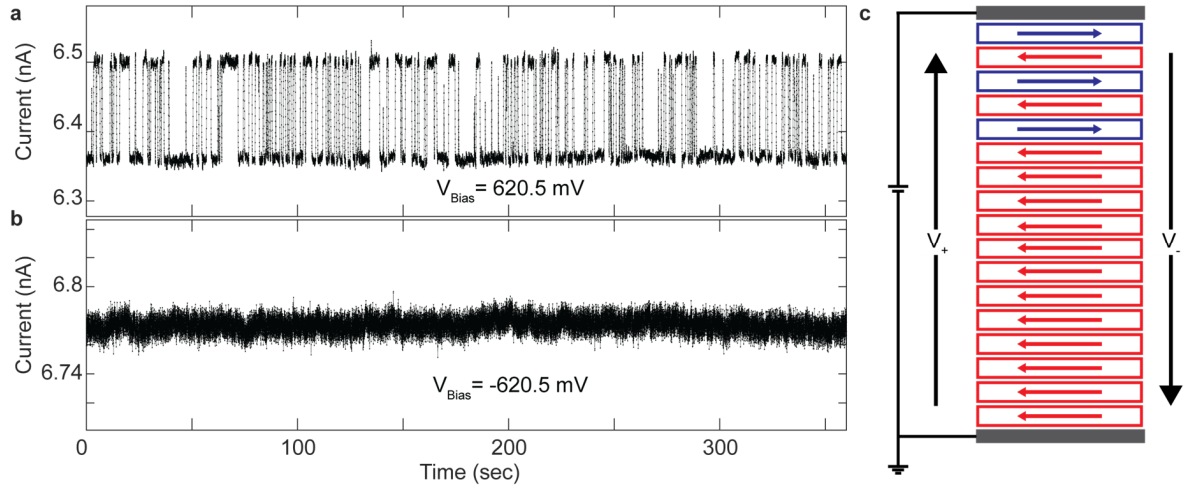

**Extended Data Fig. 5 | Bias-polarity-dependent stochastic switching indicates asymmetric vertical magnetic domain structure. a-b,** Tunneling current over time of a metastable domain with a positive (**a**) and negative (**b**) bias applied to the MTJ. Despite a similar magnitude of current, no switching is observed under negative bias, ruling out heating effects. Instead, the data is consistent with an asymmetric vertical domain structure, as illustrated in (**c**). A plausible scenario is that when a positive voltage is applied, the FM layers polarize the tunneling electrons. These spin polarized electrons apply a spin-transfer torque like effect to the AFM layers, enhancing the stochastic switching. On the other hand, when a negative bias is applied, the electrons are not highly polarized and do not exert a spin-transfer torque on the FM layers.

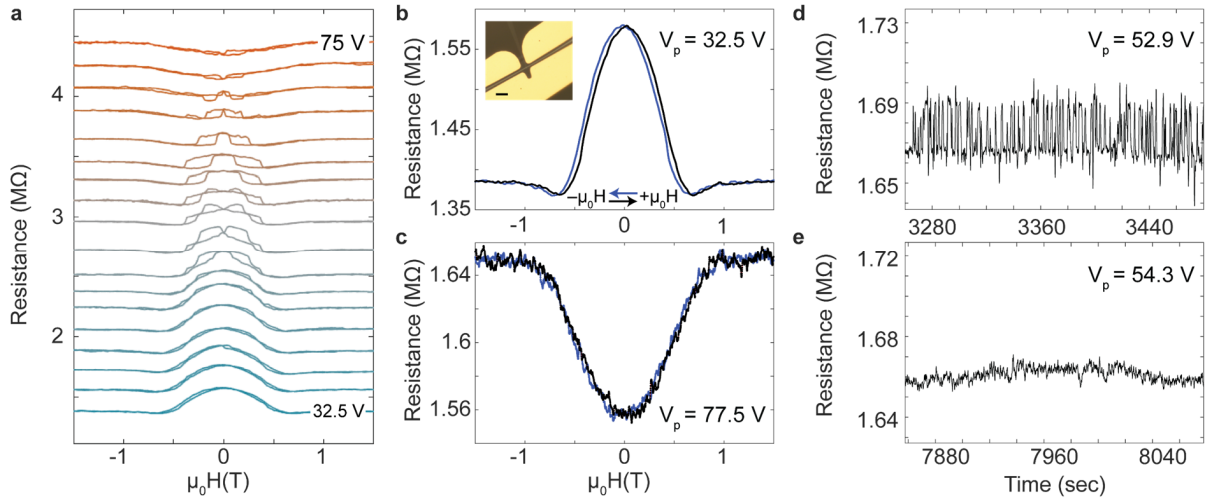

**Extended Data Fig. 6 | Strain switching in a six-layer MTJ. a,** Magnetoresistance sweeps of a MTJ with a six-layer CrSBr tunnel barrier as the piezo voltage $V_p$ is increased from 32.5 V to 75 V. The domain behavior at piezo voltages between the low-strain AFM (32.5 V) and high-strain FM (75 V) states is much simpler than the ~ 16-layer device presented in the main text, providing additional evidence that vertical, layer-dependent domains are the origin of the complex hysteretic domains behavior during the magnetic phase transition. The magnetic field is applied along the *a* axis at a temperature of 20 K. **b-c,** Magnetoresistance sweeps in the low strain AFM (**b**) and high strain FM (**c**) states, showing the characteristic switching from negative to positive MR. The optical image of the device is shown inset in **b** (scale bar 5 µm). **d-e,** Resistance over time at select piezo voltages during the magnetic phase transition. Stochastic domain switching (**d**) which can be stabilized by slightly increasing strain e) are observed. These results highlight the potential for extending the strain-programmable vdW MTJs to the 2D limit.

# Supplementary information for

# Strain-programmable van der Waals magnetic tunnel junctions

**Authors:** John Cenker[1], Dmitry Ovchinnikov[1], Harvey Yang[1], Daniel G. Chica[2], Catherine Zhu[1], Jiaqi Cai[1], Geoffrey Diederich[1,3], Zhaoyu Liu[1], Xiaoyang Zhu[2], Xavier Roy[2], Ting Cao[4], Matthew W. Daniels[5], Jiun-Haw Chu[1], Di Xiao[4,1], Xiaodong Xu[1,4,*]

[1] Department of Physics, University of Washington, Seattle, Washington 98195, USA
[2] Department of Chemistry, Columbia University, New York, NY 10027 USA
[3] Intelligence Community Postdoctoral Research Fellowship Program, University of Washington, Seattle, WA, USA
[4] Department of Materials Science and Engineering, University of Washington, Seattle, Washington 98195, USA
[5] Physical Measurement Laboratory, National Institute of Standards and Technology, Gaithersburg, MD, 20899, USA

*Correspondence to: xuxd@uw.edu

**Supplementary Text: Additional stochasticity analysis of switching data taken near ρ = 0.5**

Since the tunneling current is sampled much faster than the switching rate (~ .14 sec), switching data collected over 200 seconds was downsampled and tested using 15 tests from the NIST test suite[1]. Maurer's Universal Test was excluded since the binary sequence was not long enough. The full sampling time dependence is shown below, using a standard threshold p-value of .01. The grey line indicates the sequence passed all of the 15 considered tests. The red line indicates the average domain switching time obtained by dividing the total number of switches by the total time window.

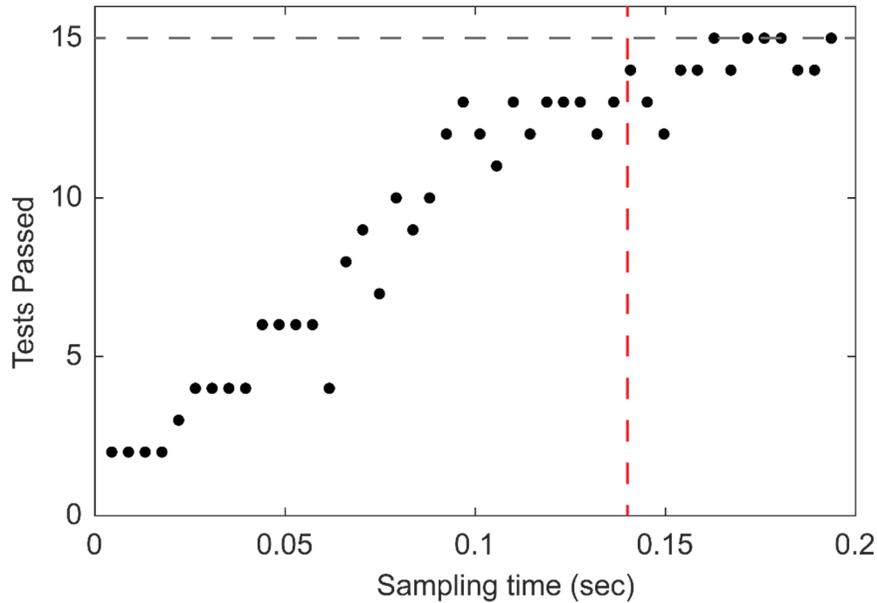

In addition to the NIST test suite, we analyzed the dwell time, i.e. the time between switches, of the 0 and 1 states. The extracted dwell times are plotted as a histogram for the 0 and 1 states, following an exponential envelope as expected for a Poisson process.

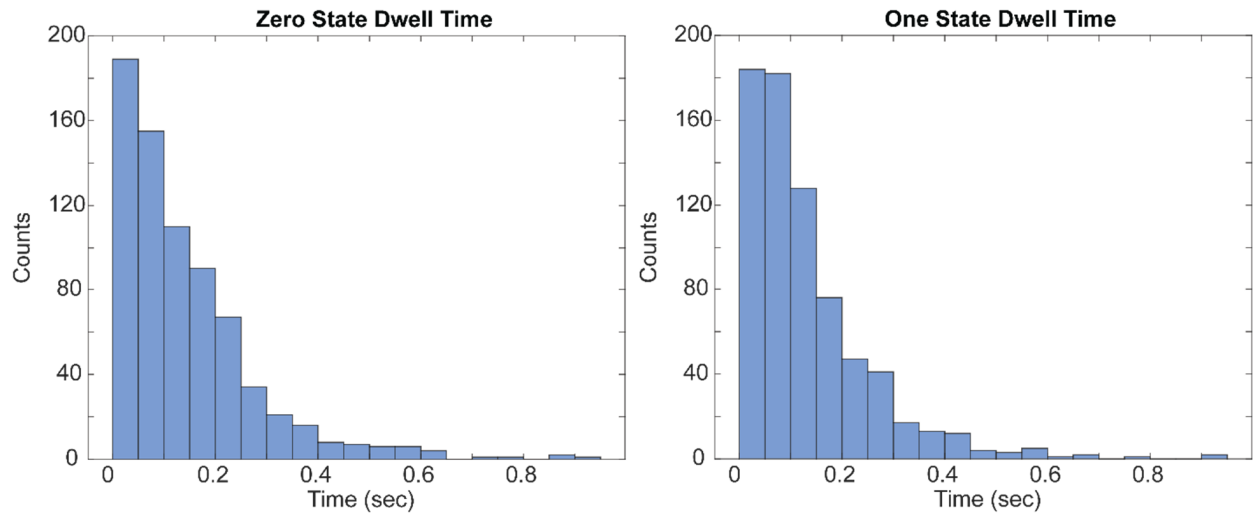

We then plot the logarithm of the histogram bin counts (N) versus the dwell time:

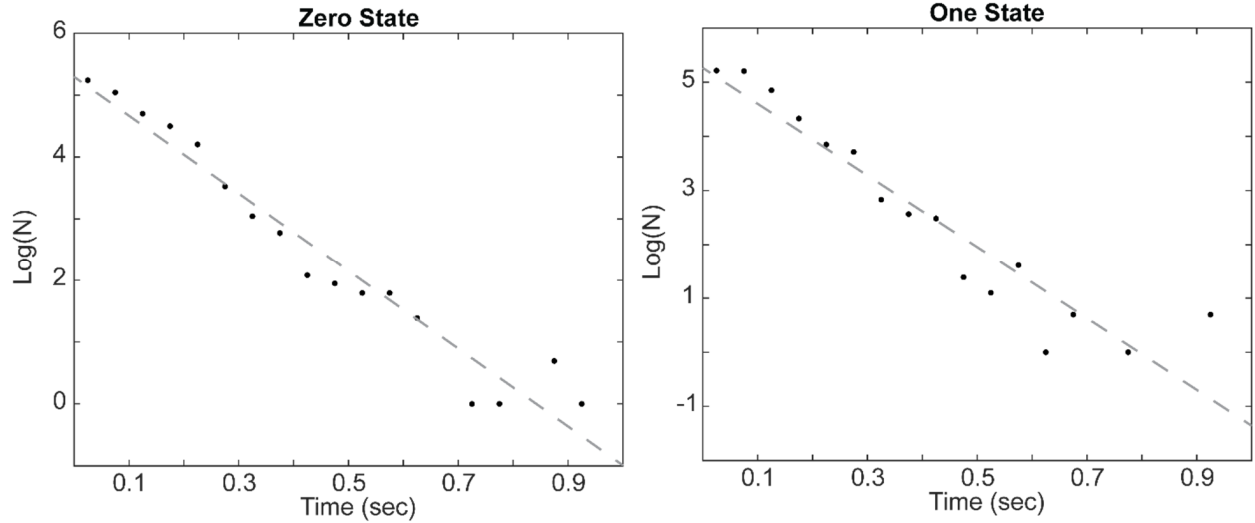

From the linear fits, we find that the characteristic lifetime, τ, of the 0 and 1 states are $\tau_0 = 159 \pm 9$ ms and $\tau_1 = 151 \pm 9$ ms, respectively, where the uncertainty is determined by the standard deviation of the linear fit. Based on this analysis and the NIST test suite results, we conclude that the strained MTJ can generate binary sequences with a high degree of randomness.